%% file: Submitted_EPL_April13.99
\documentstyle[epsfig]{EuroPhys}
\input EuroMacr

\begin{document}
%
%
%
\euro{}{}{}{}
\Date{}
\shorttitle{Structure of Colloid-Polymer Mixtures}
%
%
%
\title{The structure of colloid-polymer mixtures}
\author{A. A. Louis\inst{1}, R. Finken\inst{1}\footnote{Present address: 
Institut f\"{u}r Theoretische Physik II, Universit\"{a}t D\"{u}sseldorf,
D-40225 D\"{u}sseldorf, Germany}
J-P. Hansen\inst{1} }
\institute{
     \inst{1} University Chemical Laboratories, 
Lensfield Rd, Cambridge CB2 1EW, UK\\}
     
%
%
\rec{}{}
%
%
%
\pacs{
\Pacs{61}{20.-p}{Structure of liquids}
\Pacs{82}{70.Dd}{Colloids}
      }
\maketitle
%
%
%
\begin{abstract}
We investigate the structure of colloid-polymer mixtures by  calculating 
the structure factors for the Asakura-Oosawa model in the PY approximation. 
We discuss the role of potential range, polymer concentration
and polymer-polymer interactions on the colloid-colloid structure.
Our results compare reasonably well with the recent  experiments of 
Moussa\"{i}d {\em et.\/ al.\/}\cite{Mous99} for small wavenumber $k$, but 
we find that  the Hansen-Verlet freezing criterion is violated 
when the liquid phase becomes marginal.
\end{abstract}
%
%
%
%
%

Suspensions of sterically stabilized colloidal particles in an organic
solvent containing non-adsorbing polymer have been widely studied
experimentally and theoretically because of their remarkable phase
behaviour.  The non-adsorbing polymer coils induce an attractive
depletion interaction between the colloidal particles, the range and
depth of which can be easily ``tuned'' by varying the polymer
molecular weight (and hence its radius of gyration $R_g$) and the
polymer concentration $n_p$.
This  makes polymer-colloid mixtures an
important experimental model system to study the role of the range of 
attractive interactions on phase behaviour.   

If $R_c$ is the radius of a spherical colloidal particle, and $\xi =
R_g/R_c$ denotes the size ratio, it was predicted
theoretically\cite{Gast83,Lekk92}, and shown experimentally\cite{Ilet95}
and by simulations\cite{Meij94} that the phase diagram in the $(\eta_c,
n_p)$ plane (where $\eta_c = 4 \pi n_c R_c^3/3$ denotes the colloid packing
fraction) exhibits colloid ``gas'', ``liquid'' and ``solid'' phases, when
$\xi > \sim 0.3$, while the intermediate ``liquid'' phase becomes
``marginal'' for $\xi \approx 0.3$, and disappears completely for smaller
size ratios\footnote{There is still some discrepancy between theory and 
simulations which predict a marginal liquid for $\xi\sim 0.4$\cite{Gast83,Meij94}
, and 
experiment, which finds a marginal liquid at $\xi\sim 0.24$\cite{Ilet95}.}.
 The previous theoretical work mostly focuses on calculating
free energies and the resulting phase diagrams and no attempts were made to
investigate the structure of the colloid-polymer mixtures, as embodied in
the partial structure factors $S_{\alpha \beta}(k)$ ($1 \leq \alpha \leq
\beta \leq 2)$, where the indices $1$ and $2$ refer to the colloid and
polymer species respectively).  This is because the van-der-Waals like
mean-field theories in\cite{Gast83,Lekk92} do not readily yield structural
information, and because until very recently no experimental data were
available.  This gap has been filled by the two-colour dynamic
light-scattering measurements of Moussa\"id {\em et.\/ al.\/}\cite{Mous99}, which
provide data for the colloid-colloid structure factor, $S(k) = S_{11}(k)$,
over a limited range of wavenumbers $k$, for thermodynamic states
 at triple-point conditions, and for three values of the size
ratio $\xi$.  In
particular, they concluded that under triple point conditions the amplitude
of the first peak of the structure factor was always near $2.5$ while the
small $k$ behaviour showed a steep rise reminiscent of critical behaviour.
In other words the local fluid structure at the triple point is insensitive
to the potential range, while the large scale structure is sensitive to
this range.  These finding are analyzed here within a simple
colloid-polymer model.

In this letter we apply the familiar Percus-Yevick (PY) closure\cite{Hans86}
to the Asakura-Oosawa (AO)\cite{Asak54} model of a colloid-polymer mixture
to extract the partial pair distribution functions $g_{\alpha\beta}(r)$,
and corresponding structure factors.  The AO model treats a colloid-polymer
system as a {\em non-additive}\footnote{By non-additive we mean that the
$R_{12} \neq \frac{1}{2}(R_{11} + R_{22})$.} mixture of spheres of radius 
$R_{11}=R_c$ (the colloids)
and point particles (the weakly interacting polymer coils) which are excluded
from a sphere of radius $R_{12}=R_c+R_g$ around each colloid
\footnote{Since
interactions between polymer-coils are neglected, their activity 
$a_p =\exp\left\{ \beta \mu_p \right\}$ is 
proportional to the effective polymer concentration $n_p' = N_p/V'$, where
$V'$ is the free volume accessible to the polymer coil; $V'$ is a fraction
$\alpha V$ of the total volume due to the presence of the colloidal particles
from which the polymer coils are excluded.}. 
 This binary
system is characterized by three total and direct correlation functions, 
$h_{\alpha \beta}(r) \equiv g_{\alpha \beta}(r) -1$, and $c_{\alpha \beta}(r)$, the
Fourier transforms of which satisfy three coupled Ornstein-Zernike 
relations\cite{Hans86}:
\begin{equation}\label{eq1}
\hat{h}_{\alpha \beta}(k) = \hat{c}_{\alpha \beta} (k) + 
\sum_\gamma \rho_\gamma \hat{c}_{\alpha \gamma}(k) \hat{h}_{\gamma \beta} (k).
\end{equation}
We have solved eqs~(\ref{eq1}) numerically, subject to the PY closure relations:
\label{eq2}
\begin{eqnarray}
g_{\alpha \beta} (r) = & 0, r<R_{\alpha \beta}  
 \nonumber \\
c_{\alpha \beta} (r) = & 0, r>R_{\alpha \beta}. \label{eq2b} 
\end{eqnarray}
Note that (\ref{eq2b}) implies $c_{22}(r) =0$, since $R_{22}=0$ (point
particles).  The 2nd and 3rd OZ relations (\ref{eq1}) accordingly simplify
to yield the relations:
\begin{eqnarray}\label{eq3}
S_{12}(k) = & \sqrt{n_1 n_2} \hat{h}_{12}(k) & = \sqrt{n_1 n_2} \hat{c}_{12}(k)
S_{11}(k) \label{eq3a}\\
S_{22}(k) = & 1+ n_2 \hat{h}_{22}(k) & = 1 + n_1 n_2 \hat{c}^2_{12}(k)
S_{11}(k). \label{eq3b}
\end{eqnarray}
Note that $S_{22}(k) \neq 1$ (and hence 
$h_{22}(r) \neq 0$), despite the fact that the polymer coils do not
interact (point particles), due to correlations induced by the presence
of the colloidal particles.  Thus $g_{22}(r)$ provides a measure of the
``void structure'' of the annealed, porous ``medium'' created by the 
colloidal spheres.  

Contrary to the case of additive hard-sphere mixtures\cite{Lebo64}, the coupled
OZ/PY equations admit no analytic solution in the present non-additive
case\cite{Lebo71}; an analytic solution for the AO model only exists
in the physically irrelevant case of {\em negative} non-additivity
(which would amount to $R_g < 0$!)~\cite{Roni84}.  Accurate numerical solutions
are, however, readily obtained and results are reported below, and
compared to the experimental data and to the PY results for the {\em
effective} one-component system of colloidal particles interacting via
the AO depletion potential\cite{Asak54}.  The effective depletion
interaction follows from the gain in free volume, accessible to the
non-interacting polymer coils, due to overlaps of exclusion spheres of
radius $R_{12}$ for any configuration of the non-overlapping colloid
spheres of radius $R_{11} < R_{12}$.  The AO pair potential follows from
a second-virial-like approximation whereby only pair-wise overlaps are
considered, thus neglecting more than two-body effective interactions
resulting from overlap of three or more exclusion spheres.  This is
strictly valid only for $\xi < \xi_0=2/\sqrt{3}-1$.  For $\xi > \xi_0$
the two-component AO model can no longer be exactly mapped onto an
effective one-component system of colloidal particles with strictly
pairwise interactions.  

For a one-component system the long wavelength ($k \rightarrow 0$) 
of the structure factor $S(k)$
is given by the familiar relation\cite{Hans86}:
\begin{equation}\label{eq4} 
\lim_{k \to 0} S(k) = n_c k_B T \chi_T.
\end{equation} 
where $\chi_T$ is the isothermal compressibility of the
effective one-component system.  The situation is more complicated in
the two-component case, where the ($k \to 0$) limit of the
$\hat{c}_{\alpha \beta}(k)$ and of the $S_{\alpha \beta}(k)$ are
governed by the derivatives of the chemical potentials of the two
species, $\mu_{\alpha \beta} = n_\beta (
\partial(\mu_\alpha/k_BT)/\partial n_\beta)_{T,n_\alpha}$.  For the AO
model in the PY approximation,  where $\hat{c}_{22}(0) =1-\mu_{22} \equiv 0$,
 implying $\mu_{22} = 1$,
one finds in particular that:  
\begin{equation}\label{eq5a} \lim_{k \to 0} S_{11}(k) =
\frac{1}{\mu_{11} - \mu_{12} \mu_{21}}, \end{equation} whereas (with $n
= n_1 + n_2$ and $x_{\alpha} = n_\alpha/n$)
\begin{eqnarray}\label{eq5b} n k_B T \chi_T &= &\lim_{k \to 0}
\frac{1}{1 - x_1^2 n \hat{c}_{11}(k) - 2 x_1 x_2 n \hat{c}_{12}(k) 
-x_2^2 n \hat{c}_{22}(k)} 
\nonumber \\
& = &
\frac{1}{x_2 + x_1(\mu_{11} + 2 \mu_{12})}, \end{eqnarray}
showing that the long wavelength limit of the
colloid-colloid structure factor is not directly proportional to the
osmotic compressibility of the colloid-polymer mixture.  Note that the
PY closure implies that $\mu_{22} =1$, which is tantamount to assuming that the free
volume $V'$ is unaffected by the presence of the non-interacting
polymer-coils, {\it i.e.\/}  is independent of $n_2$; this is precisely the
basic assumption underlying the mean field theories of the phase
diagram\cite{Gast83,Lekk92}.

\begin{figure}
\begin{minipage}{69mm}
\centerline{\includegraphics[width=73mm]{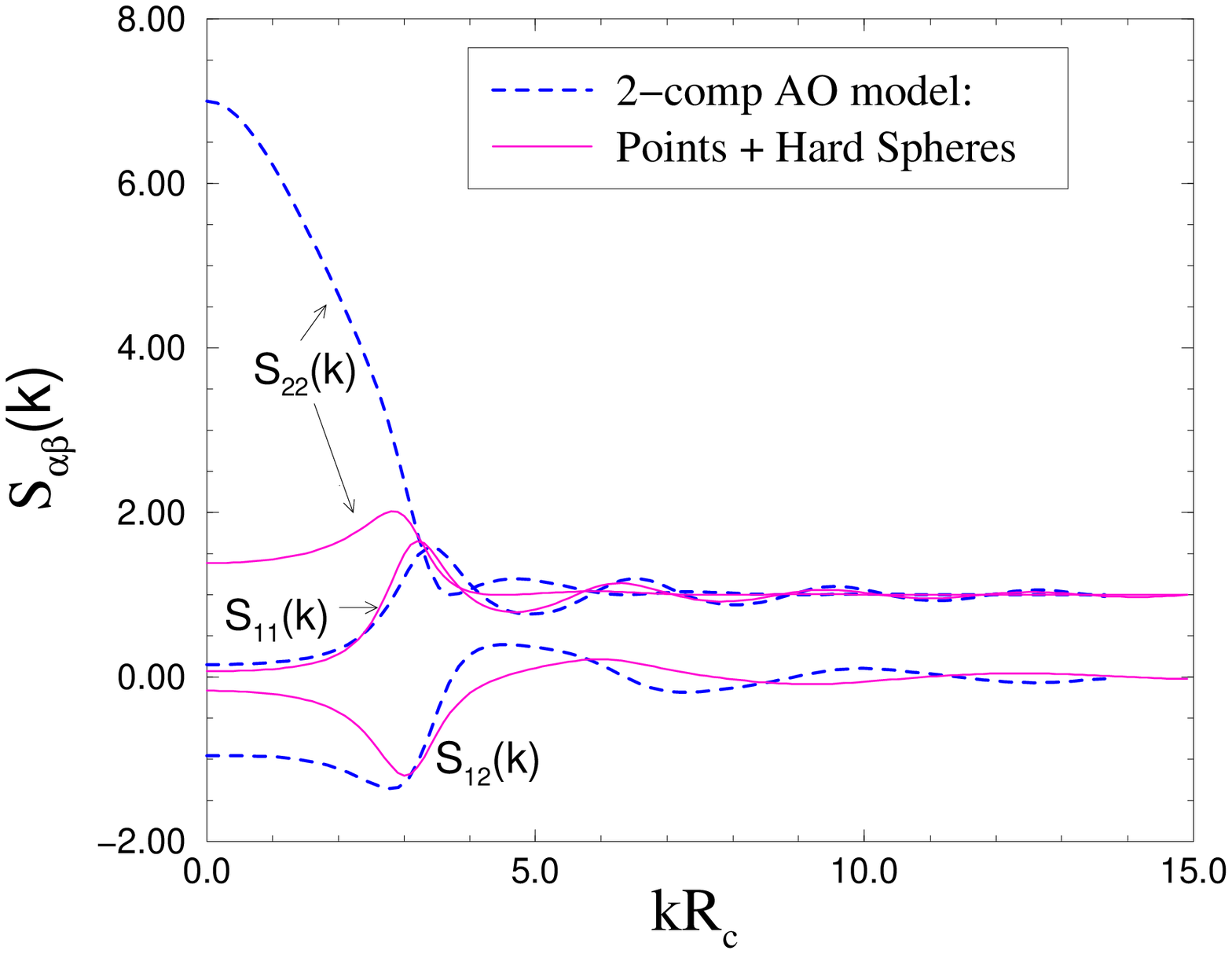}}
\end{minipage}
\hspace*{2mm}
\begin{minipage}{69mm}
\centerline{\includegraphics[width=73mm]{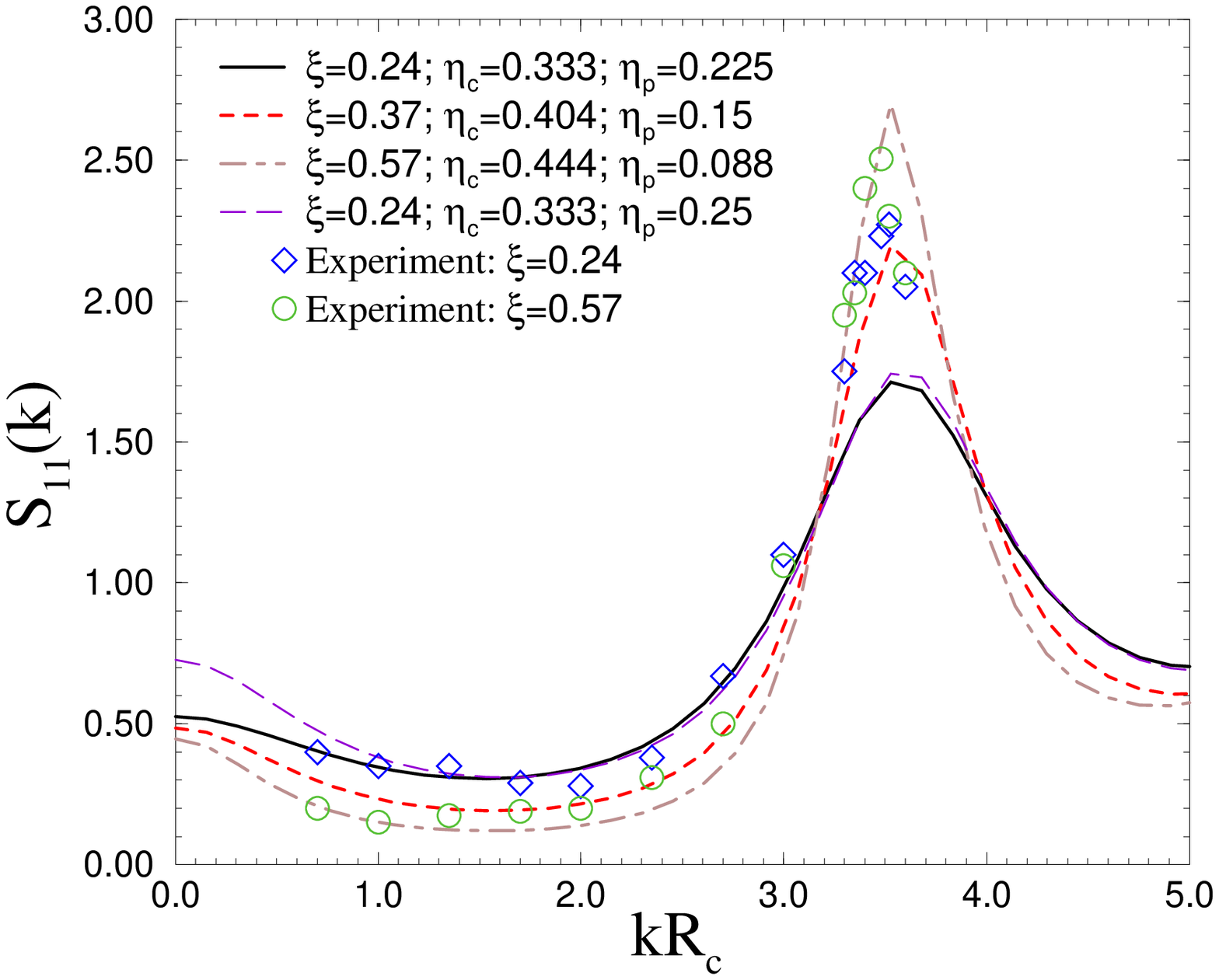}}
\end{minipage}
\caption{ Comparison of the structure factors of the 2-component AO 
model and an additive mixture of hard spheres and point particles,
calculated in the  PY approximation for $\eta_c = 0.333$, $\eta_p = 0.1$
and $\xi=0.24$.    
}
\label{Fig1}
\caption{Comparison of the experimental colloid-colloid structure factors
 of Mousa\"{i}d {\em et.\/al.\/}\protect\cite{Mous99} with the AO model in the 
 PY approximation.
 The values of the colloid packing fraction, $\eta_c$, and the size
 ratio $\xi$ are taken from experiment.  The polymer packing fraction,
 $\eta_p$, is fit to the experimental value of $S_{11}(kR_c)$ at the 
 smallest accessible wavenumber ($kR_c=0.7$).
 The resulting polymer packing fractions are $\eta_p = 0.225, 0.15$ and $0.089$
 respectively, compared to the measured values of $\eta_p = 0.13, 0.13$ and $0.1$
 respectively. The 2nd theoretical line at $\xi=0.24$ demonstrates the
 insensitivity  of the 1st peak  versus the sensitivity of the low $k$ 
 behaviour of the colloid-colloid structure factor, $S_{11}(k)$, to changes in $\eta_p$.}
 \label{Fig2}
\end{figure}

The overall behaviour of the three partial structure factors is illustrated in
fig.~\ref{Fig1}, where a comparison is made with the analytic PY results for an
additive mixture of spheres and point-particles (corresponding to the $R_g =0$
limit of the AO model; $S_{11}(k)$ is then unaffected by the point particles and thus
equivalent to a one-component hard-sphere model).  The non-additivity is seen to
strongly affect $S_{12}(k)$ and $S_{22}(k)$, while the colloid-colloid structure
factor, $S_{11}(k)$, is less sensitive.  The greater sensitivity of $S_{12}(k)$
and $S_{22}(k)$ may be traced back to the significant change in void structure
when the excluded volume radius $R_{12}$ increases (non-additivity).

Fig.~\ref{Fig2} compares $S_{11}(k)$ calculated for the AO  model to
the experimental data of Moussa\"{i}d {\em et.\/al.\/}\cite{Mous99} for the 
three
size-ratio's they used:  $\xi = 0.57$, which corresponds to a situation
where the critical and triple points are well separated, $\xi = 0.37$,
where the critical point and triple point almost merge and the liquid is
``marginal'', and $\xi = 0.24$, where there is no stable liquid region and
the critical point becomes metastable w.r.t.  the melting line (Note that
in the experiment, $\xi=0.24$ (and not $\xi=0.37$) corresponds to the marginal
liquid.).  The
 colloidal packing fractions, $\eta_c$, and size-ratios, $\xi$,
were taken from the published experimental values while the polymer packing
fractions, $\eta_p = 4 \pi n_p R_g^3/3$, (which depend on $n_p$ {\em and}
$\xi$) were adjusted to fit the experimental structure factors, $S_{11}(k)$, at
$k=0.7/R_c$\footnote{We note that taking the experimental polymer number
density as fixed and varying the effective size-ratio parameter, $\xi$,
would be more consistent with fitting to experiment since it is not clear
that $\xi$ should be exactly equal to  $R_g/R_c$.  For example we find
we can fit the small $k$ behaviour of $S_{11}(k)$ at the experimental
polymer density for the shortest polymers by changing $\xi$ from $0.24$ to
$0.28$. This seems to be consistent with recent findings on the interactions
between a gaussian polymer and a hard wall\cite{Eise97}, but is inconsistent
with another computer simulation study of the AO model\cite{Meij94}. }.  Keeping in mind the fact that PY theory tends to overestimate
the amplitude of the main peak, the agreement for $\xi = 0.57$ is
satisfactory, but  for $\xi = 0.37$ and $\xi =
0.24$ the amplitude of the main peak drops considerably, in contrast to
the experiment where the main peak is insensitive to the size ratio $\xi$.

 For a given colloidal packing fraction, $\eta_c$,
the amplitude of the first peak of the colloid-colloid structure factor is
rather insensitive to varying $\eta_p$ (or $\xi$).  While shortening the
range of the attractive interactions widens the melting transition,
bringing the liquidus line to lower colloidal packing fractions, we find
that the amplitude of the first peak of the colloid-colloid structure
factor is determined primarily by the colloid packing fraction, $\eta_c$,
and {\em not} by the proximity of the freezing transition.  In contrast,
the small $k$ behaviour of the structure factor is strongly affected by the
proximity of a critical point or a spinodal line.  These conclusions are
consistent with recent simulations of one-component systems with
short-range potentials\cite{Dijk99,Lekk99}, but inconsistent with 
the experiments\cite{Mous99}.

\begin{figure}
\begin{center}
\begin{minipage}{69mm}
\centerline{\includegraphics[width=73mm]{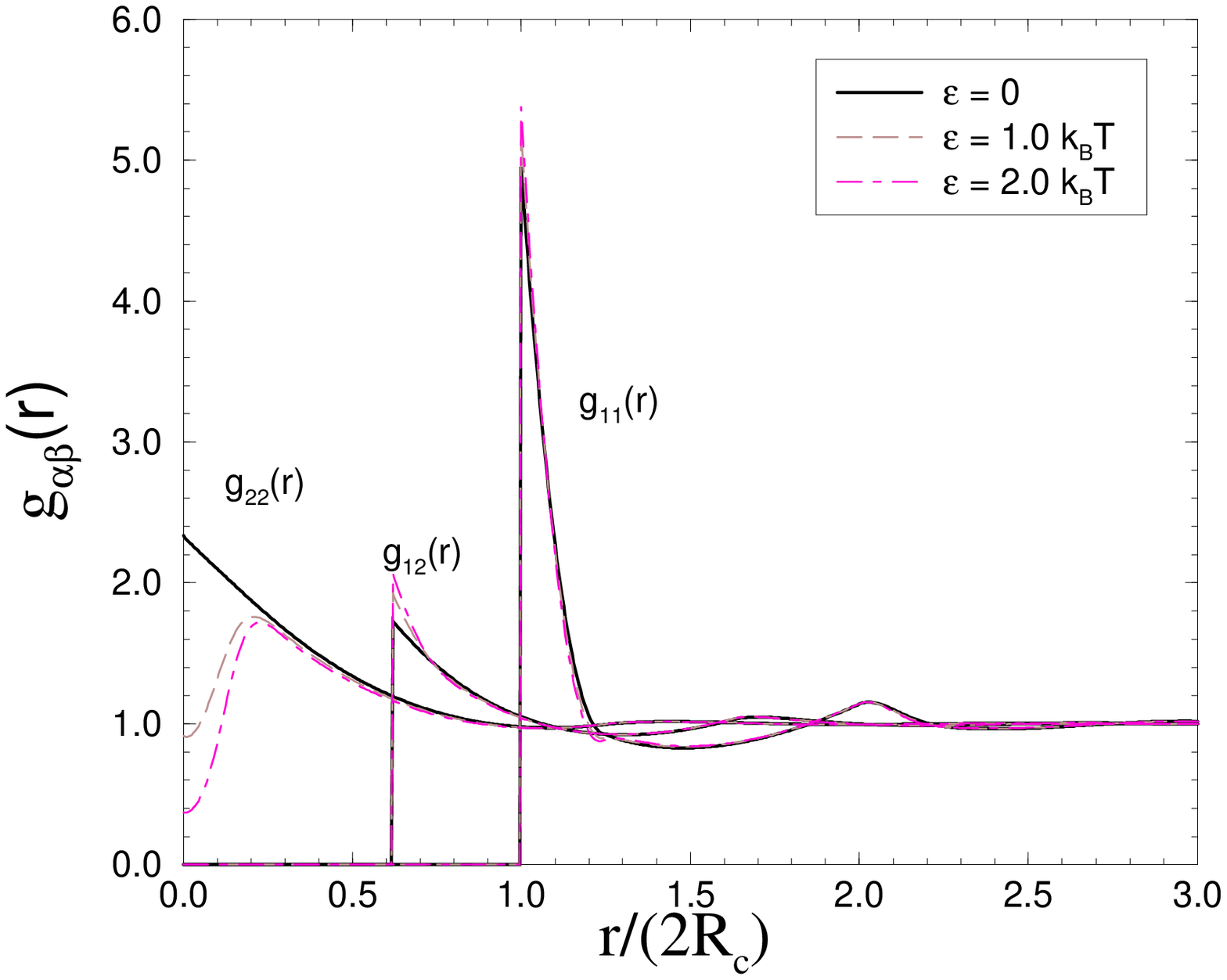}}
\end{minipage}\hspace*{2mm}
\begin{minipage}{69mm}
\centerline{\includegraphics[width=73mm]{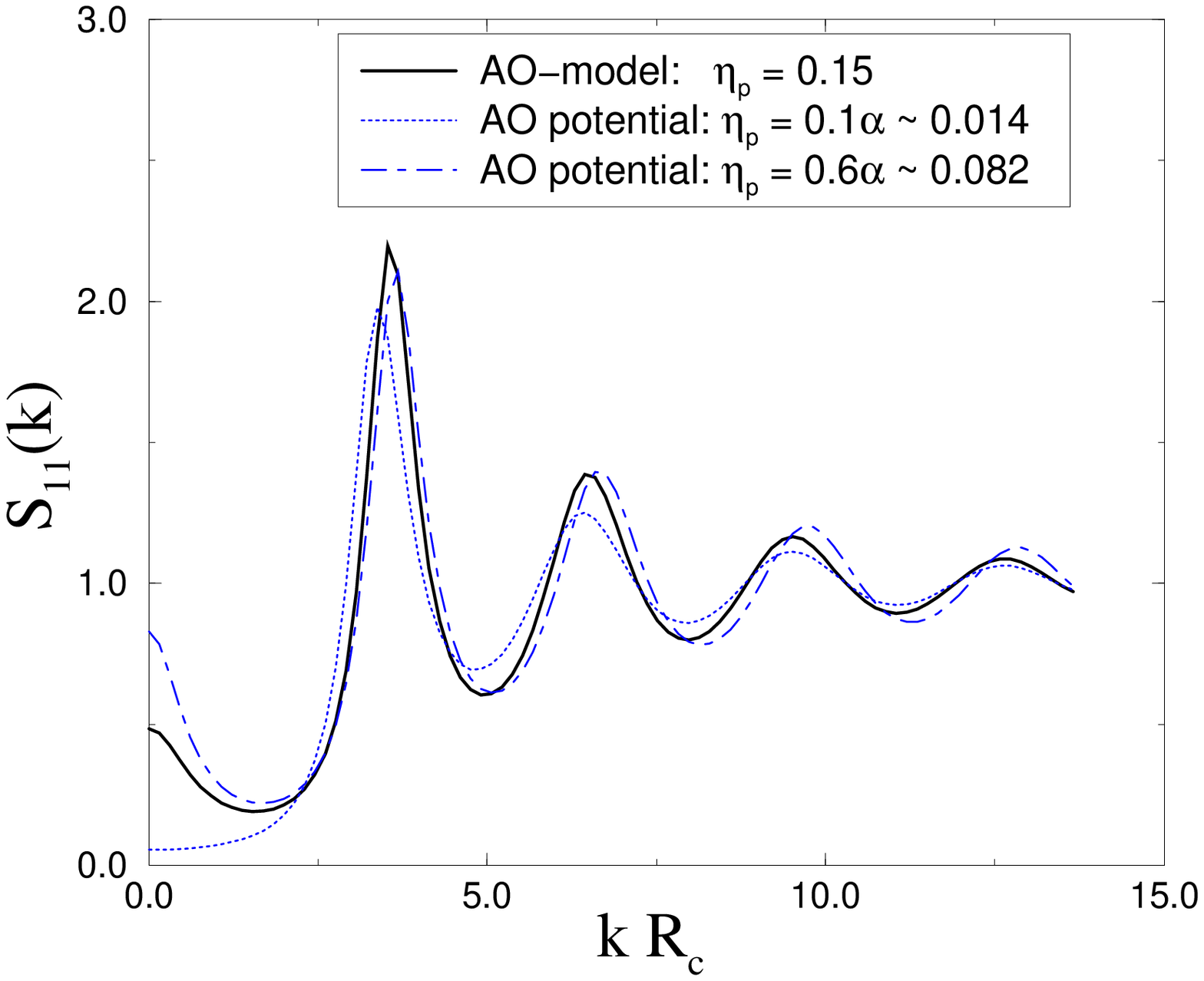}}
\end{minipage}
\caption{Pair distribution functions for a colloid-polymer mixture with 
repulsive polymer-polymer interactions of the form of 
equation\protect~(\ref{eq6}) with $\eta_c=0.333$, $\eta_p=0.13$ and $\xi=0.24$.
 While the polymer-polymer radial distribution
function $g_{22}(r)$ is strongly affected, the colloid-colloid radial
distribution function $g_{11}(r)$ is very weakly affected by interactions.  }
\label{Fig3}
\caption{Comparison of the colloid-colloid structure factor for the 
 AO model and the AO potential for $\xi=0.37$ ,
$\eta_c = 0.404$ and  different values of the polymer packing fraction $\eta_p$.
The AO potential polymer fugacities are converted to polymer packing fractions
through  the scaled-particle free-volume factor
$\alpha(\eta_c,\xi)$.} \label{Fig4}
\end{center}
\end{figure}

Perhaps the most notable approximation in the AO model is the complete
neglect of polymer-polymer interactions\cite{Warr95,Schw98}.  To obtain a
rough estimate of the influence of such interactions on the pair structure
of a colloid-polymer mixture we use the following model for an effective
polymer-polymer pair potential: 
 \begin{equation}\label{eq6} v_{22}(r) =
\epsilon \exp (- r/2R_g)^2, \end{equation} 
where $r$ is the distance between the
centers of mass of the two polymer coils, and $\epsilon$ is an energy scale
which is expected to be of order $k_BT$, since the effective interaction is
largely of entropic origin. This form of the effective  potential
is consistent with the findings  of direct simulations
of hard-sphere polymer chains\cite{Daut94}. Eq~(\ref{eq6}) is a simple parameterization of
the free energy cost linked to the interpenetration of two polymer coils;
the range of such an effective interaction is obviously of the order of 
 the radius of gyration, and depending on the solvent-monomer and
monomer-monomer interactions, $\epsilon$ could be positive or negative.  In
a good solvent $\epsilon$ is expected to be positive (effective repulsion).
It is easy to incorporate the pair potential~(\ref{eq6}) in the PY theory
eq.~(\ref{eq2b}). The effect of 
turning on the polymer-polymer interactions  is illustrated by 
the radial-distribution
functions in fig. \ref{Fig3}.
The interaction (\ref{eq6}), 
turns out to have
a weak influence on the colloid-colloid structure, 
 while, as expected,
the polymer-polymer correlations, which are only indirect in the point 
particle case, are strongly affected by $v_{22}(r)$.    Thus it appears
that the discrepancy between the AO model and the experiments does not 
lie in the neglect of polymer-polymer interactions.

A final question to address is the role of many-body interactions, naturally
taken into account in the two-component system, but neglected in the popular
one-component approximation to the AO model:  the hard-sphere system with an
attractive AO potential.  The comparison is somewhat complicated by
thermodynamics, since the strength of the AO potential is directly
proportional to the polymer activity $a_p$, while the AO model is most
naturally interpreted in terms of the polymer density $n_p$ (or polymer
packing fraction $\eta_p$).  Since the
polymers are non-interacting, however, the two can be directly related
 through the
free-volume fraction $V'/V=\alpha$.
In fig.~\ref{Fig4} we compare the PY 
results for the 2-component AO model in the marginal liquid state 
($\xi = 0.37$)
to the 1-component AO potential with the $\eta_p$ scaled by the 
appropriate free-volume fraction $\alpha$ for which we take the
scaled-particle approximation, which has been shown to be
sufficiently accurate for the qualitative behaviour we are 
addressing\footnote{The overall free-volume is well represented by
the scaled-particle theory expression for $\alpha$\cite{Meij94}, but the 
thermodynamics of theories based on free volume, such as those in
{\em ref} \cite{Lekk92}, depend critically on $d^2\alpha/dn_c^2$, for which the 
scaled-particle approximation is less accurate\cite{Dijk99}.}. 
 Significant differences between the structure
factors calculated within the one and two-component models are observed for
identical values of $\eta_p$, particularly at  small $k$. Use of the AO
potential will lead to a severe {\em underestimate} of  the polymer density,
$n_p$; the discrepancy
becomes progressively worse as $\xi$ increases. For example, for $\xi=0.57$
the low-k fit to the experimental $S(k)$ yields $\eta_p=0.088$ for the AO 
{\em model} but 
$\eta_p \sim 0.01$ for the AO {\em potential}.  We note that the rise 
in the structure factor
of the AO potential system is consistent with the location of 
the spinodal line calculated in the mean-field theories\cite{Gast83,Lekk92},
while the rise in the structure factor of the AO model is more consistent
with the experiments.  
 Since the same
PY level of approximation is used to calculate both structure factors, the
disagreement  suggests the inadequacy of the AO potential or the
assumption of pair-wise additivity in the reduction to an effective
one-component system.  It must be stressed however, that the PY closure
is far from exact, and may be inadequate to cope with a short-range
attractive well, as in the AO potential, particularly for the smaller 
values of $\xi$.  On the other hand PY is expected to be more accurate for
purely repulsive interactions as in the 2-component AO model.
 We are presently examining this point using thermodynamically
self-consistent closures\cite{Cacc95}.  

As
pointed out earlier, the $k\rightarrow 0$ limits of the one-component
colloid-colloid structure factor $S(k)$, and its two-component equivalent
$S_{11}(k)$ do not have the same thermodynamic meaning.    For $\eta_c = 0.404$,
$\eta_p = 0.15$ and $\xi=0.37$, the osmotic compressibility derived from 
eq.~(\ref{eq4})
for the effective one-component system ($S(k)=S_{11}(k)$), 
and from eq.~(\ref{eq5b}) for the
original two-component AO model are $n_c k_BT\chi_T = 0.49$ and 
$n_c k_BT\chi_T = 0.035$
respectively.

We may draw the following main conclusions from the present work:

a) In contrast to the experiments of Moussa\"{i}d {\em et.\/al.\/}\cite{Mous99}
we find (both with the 2-component AO model {\em and} the 1-component AO
potential) that the first peak of the colloid-colloid structure factor and the
related local fluid order is primarily determined by the colloid volume
fraction, and not by the the proximity to the melting transition.  Thus the
Hansen-Verlet criterion, which states that the first peak of the structure
factor at freezing is near $2.8$, is violated as the range of the potential is
shortened to the point where the triple point vanishes. This discrepancy
with experiment may be related to novel effects arising from an interplay
between the triple and critical points, which for $\xi=0.24$ are in 
close proximity (marginal liquid).

  b) In agreement with the experiments\cite{Mous99}, we find that the 
  large scale fluid structure and related small $k$ limit of the structure 
  factor is strongly affected by the proximity of a spinodal line or
  a critical point, and thus depends significantly on the polymer density.
  The small $k$ limit of the structure factors is in fact well described 
  by the AO model.
  
 c) Polymer-polymer interactions appear to have little impact on the
 colloid-colloid pair structure, at least within the naive ansatz of 
 eq.~(\ref{eq6}).  
  
d) The  one-component description, based on the effective AO pair potential,
 appears to be insufficient to yield a quantitatively accurate description 
 of the colloid-colloid structure factor, particularly at small $k$.  It is
 consistent with the phase-diagrams calculated in the mean-field theories,
 and similarly 
 underestimates the polymer packing fraction for larger size ratios.

 To ensure consistency of the theory, we are presently extending the PY work
 to other, thermodynamically consistent, integral equations which we will
 use to determine the phase diagram of the two-component model.  We are also
 seeking a first principles approach to the effective polymer-polymer and 
 polymer-colloid interactions.

%
\stars
 The authors are grateful to Wilson Poon for suggesting the PY analysis, and
 for providing a pre-print of ref\cite{Mous99}, and  Patrick Warren
 for useful suggestions.  AAL would like to thank the  EC  for support 
 through the grant EBRFMBICT972464.

%
%

\end{document}

%% file: EuroMacr.tex
\def\And{{\rm and\ }}

\def\stars{\bigskip\centerline{***}\medskip}

\newif\ifboo \boofalse

\def\Review#1{\boofalse{\it #1},}
\def\Name#1{{\sc #1},}
\def\Vol#1{\ifboo Vol. {\bf #1}\else{\bf #1}\fi}
\def\Year#1{\ifboo #1\else(#1)\fi}
\def\Book#1{\bootrue{\it #1},}
\def\Page#1{\ifboo {\rm p. #1}\else{\rm #1}\fi}